\documentclass[aps,prb,twocolumn,superscriptaddress,a4paper,floatfix,amsfonts,amsmath,amssymb,citeautoscript]{revtex4}

\usepackage{mathptmx}
\usepackage{float}
\usepackage[scaled=0.9]{helvet}
\usepackage[utf8]{inputenc}
\usepackage{graphicx}
\usepackage[pdftex,dvipsnames]{xcolor}
\usepackage[pdftex]{hyperref}
\hypersetup{
    pdftitle={},
    colorlinks,
    citecolor=blue
    }
\usepackage{upgreek}
\usepackage{bm}
\usepackage{textcomp}
\usepackage{xspace}
\usepackage{xargs}
\usepackage{verbatim}
\usepackage[normalem]{ulem}
\usepackage{ifthen}
\usepackage{enumitem}

\newcommand{\affiliationPDI}{\affiliation{Paul-Drude-Institut f\"ur Festk\"orperelektronik, Leibniz-Institut im Forschungsverbund Berlin e.V.,
Hausvogteiplatz 5-7, 10117 Berlin, Germany}}

\newcommand{\fig}[2]{Fig.~\ref{#1}\ifthenelse{\equal{#2}{}}{}{(\lowercase{#2})}}
\newcommand{\eq}[1]{Eq.~\eqref{#1}}

\newcommand{\sect}[1]{Sec.~\ref{#1}}

\newcommand{\omegac}{\ensuremath{\omega_\text{c}}}
\newcommand{\kb}{\ensuremath{k_\text{B}}}
\newcommand{\vh}{\ensuremath{V_\text{H}}}
\newcommand{\ns}{\ensuremath{n_\text{s}}}
\newcommand{\rh}{\ensuremath{R_\text{H}}}
\newcommand{\mB}{\ensuremath{\mu_\text{B}}}
\newcommand{\vv}{\ensuremath{{\upsilon}}}
\newcommand{\mus}{\ensuremath{\mu_\text{S}}}
\newcommand{\mud}{\ensuremath{\mu_\text{D}}}
\newcommand{\jp}{\ensuremath{j_\text{pers}}}
\newcommand{\ji}{\ensuremath{j_\text{impo}}}
\newcommand{\Ip}{\ensuremath{I_\text{pers}}}
\newcommand{\Ii}{\ensuremath{I_\text{impo}}}
\newcommand{\ve}{\ensuremath{\varepsilon}}
\newcommand{\nl}{\ensuremath{n_\text{LL}}}

\graphicspath{{./}{./figures/}}

\begin{document}

\title{Current density distribution for the quantum Hall effect}

\author{Serkan Sirt}
\affiliationPDI
\author{Stefan Ludwig}
\email{ludwig@pdi-berlin.de}
\affiliationPDI

\begin{abstract}
Our microscopic understanding of the integer quantum Hall effect is still incomplete. For decades, there has been a controversial discussion about ``where the current flows'' if the Hall resistance is quantized. Here, we analyze the current density distribution in a Hall bar based on the screening properties of a two-dimensional electron system in the quantum Hall regime. Compared to previous publications, we consider the non-local nature of the quantum Hall effect and include a related persistent current that exists inside a Hall bar if the Hall resistance is quantized. We find, that the persistent current density decreases with increasing Hall voltage. Accounting for this dependence, we find, that the current flows in the opposite directions along opposite edges of the Hall bar, while the share of the current imposed  through the contacts flows along the edge of the Hall bar with the higher chemical potential.
\end{abstract}

\maketitle

\section{Introduction}\label{sec:introduction}
The integer quantum Hall effect (QHE) is one of the most fundamental and important phenomena in condensed matter physics. The quantized Hall resistance serves as a reference for the SI (Syst\`eme International) unit system \cite{si-brochure2019,Delahaye2003}, while new applications based on the QHE are likely to emerge in the context of quantum technologies. To realize such applications, it is important to understand the microscopic distributions of the electrostatic potential as well as the current density. For instance, the electrostatic details influence the stability of the quantized Hall plateaus, while the current density distribution also affects the quantum mechanical phase of the charge carriers \cite{Ofek2009}. For a known confinement potential of a hypothetical Hall bar without disorder, the distribution of the electrostatic potential can be numerically predicted with a high accuracy \cite{Gerhardts2008}. Nevertheless, the question of how to infer from the electrostatics to the microscopic distribution of the current in the case of the QHE is still the subject of controversal debate. Clarity will not only pave the way for new applications but also enhance our understanding of the  fractional, spin or anomalous quantum Hall effects.

The text book explanation of the QHE, namely the Landauer-Büttiker picture, suggests that for the plateaus of the QHE all current flows inside one-dimensional channels along the edges of the Hall bar \cite{Buettiker1986,Buettiker1988}. This view is challanged by experiments \cite{Fontein1991,Yacoby1999,McCormick1999,Weitz2000} and theoretical calculations taking into account the Coulomb screening of the charge carriers \cite{Guven2003,Siddiki2003,Siddiki2004,Gerhardts2008}. Scanning probe experiments \cite{Weitz2000} and the so-called screening theory \cite{Gerhardts2008} suggest that near the low magnetic field end of a plateau the entire Hall potential drops within narrow strips near the edges of the sample. However, as the magnetic field is increased these strips of finite potential gradient widen and extend into the Hall bar, until ---near the high magnetic field end of the plateau--- the Hall potential drops continuously across the entire Hall bar. If the local current density is linked to the gradient of the potential, the scanning probe experiments indicate a continues transition from edge to bulk currents as the magnetic field is increased for each plateau of the QHE \cite{Guven2003}.

A direct non-invasive measurement of the current density distribution in the regime of the QHE is impossible. Nevertheless, carefully designed transport experiments allow one to study important aspects of the current flow. For example, by using a miniature ohmic contact in the center of a high mobility Hall bar as a current probe, recently the predicted transition from edge to bulk current was observed while the magnetic field was increased along the plateaus \cite{Sirt2025a}. In another experiment \cite{Sirt2025b}, multiterminal measurements were used to test whether current flows unidirectionally across the entire Hall bar or follows a chiral trajectory along the sample edge. The measurements suggest that for the entire plateaus the current follows a chiral trajectory, while in between plateaus, the current flows unidirectional as described by the Drude model \cite{Sirt2025b}.

In summary, the electrostatic potential distribution predicted by the screening theory confirms for the plateaus of the QHE the results of scanning probe measurements \cite{Weitz2000}. Transport experiments corroborate that for the plateaus of the QHE the current flows where the potential gradient is finite. This gives rise to chiral current flow, including both, edge and bulk currents depending on the magnetic field \cite{Sirt2025a,Sirt2025b}.

In the present article, we present a simple model for the current density distribution, which consistently explains these experimental findings. For simplicity, we consider a two-terminal sample consisting of a two-dimensional electron system (2DES) connected at both ends to ohmic contacts. A current \Ii\ is imposed via these contacts and flows through the 2DES. Poor ohmic contacts or additional electrically floating contacts, typically used as voltage probes, can influence the stability of the QHE but do not change the general physics discussed here. We avoid such implications, by restricting our analysis to the plateau regions of the QHE.

\subsection*{Non-local nature of the QHE}

The ballistic trajectory of a charged particle exposed to a magnetic field is chiral, a fundamental property related with the axial symmetry of the magnetic field vector. In solids, however, boundary conditions and momentum scattering can alter the impact of the axial symmetry and obscure the chirality of the trajectories of carriers. This is clearly the case for the classical Hall effect described by the Drude model, as it predicts a homogeneous unidirectional current density. The Drude model uses a relaxation time ansatz, which is local in nature and well justified in the diffusive regime for $\mu B\lesssim1$ \cite{Sirt2025b}, where $\mu$ is the mobility and $B$ the magnitude of the magnetic field component perpendicular to the plane of the 2DES.

The situation is different in the regime of the QHE with $\mu B\gg1$ (and $\kb T\ll\hbar\omegac$ with the Boltzman constant \kb, $\hbar=h/2\pi$, the Planck constant $h$ and the cyclotron frequency \omegac), where the current density is governed by the Landau quantization \cite{vonKlitzing1980}. In the 2DES of a Hall bar, for the plateaus of a quantized Hall resistance a local energy gap gives rise to regions of suppressed carrier scattering and Coulomb screening, such that within these regions current can flow virtually without dissipation \cite{Chklovskii1993a}.

An externally applied current \Ii\ generates the Hall voltage, $\vh=\rh\Ii$, dropping between the sample edges, where $\rh=h/(e^2\nu)$ is the Hall resistance and $\nu$ the local filling fraction of the Landau levels (LLs). According to the screening theory, $\nu$ remains integer within the extended regions of suppressed Coulomb screening, such that throughout the plateaus of the integer QHE \Ii\ flows dissipationless through the entire Hall bar. The dissipation related with the Hall resistance happens within ``hot-spots'' near the current carrying contacts \cite{Klass1991,Komiyama2006}.

The ``hot-spots'' and the dissipationless current within the Hall bar are signatures of a non-local nature of the QHE.

\subsection*{Persistent current of the QHE}

The non-local nature of the QHE has a surprising consequence, namely a macroscopic dissipationless current density that persists even in equilibrium, i.e., for $\Ii=0$ \cite{Chklovskii1993a,Wexler1994,Geller1994,Geller1995,Komiyama1996,Guven2003}. While this additional current is often labeled ``equilibrium current'', here we call it persistent current, \Ip. Note, that this macroscopic \Ip\ is not related with the persistent current in mesoscopic rings \cite{Geven1984}. Instead, it is driven by the effective electric fields corresponding to the gradient of the confinement potential near the edges of a Hall bar. Such local electric fields exist due to the above mentioned energy gap between LLs, which effectively prevents screening.

A previous calculation of \Ii\ in Rev.\ \onlinecite{Guven2003} applies a local version of Ohm's law to the distribution of the chemical potential predicted by the screening theory \cite{Chklovskii1992,Chklovskii1993a,Fogler1994,Lier1994,Siddiki2003,Siddiki2004,Gerhardts2008} and thereby completely neglects \Ip. It predicts that \Ii\ generally flows in the same direction along both edges of the Hall bar. However, in this article, we show, that  \Ip\ depends on \Ii, such that for the case of a finite \Ii, the neglect of \Ip\ does not yield the correct current density distribution.

Our analysis presented in this article builds on the predictions of the screening theory, in common with Ref.\ \onlinecite{Guven2003}. However, we appretiate the non-local nature of the QHE and explicitly include \Ip\ in our model. Driven by the gradient of the confinement potential near the sample edges, \Ip\ flows in opposite directions along the opposing egdes of a Hall bar. Thereby, \Ip\ is carried by electrons at energies below the chemical potentials of both leads. Clearly, it cannot contribute to the imposed current, \Ii\ (carried by electrons at energies between the chemical potentials of both contacts). However, we find that \Ip\ decreases as \Ii\ is increased. Accounting for this dependence, $\Ip(\Ii)$, we determine the distributions of \Ip\ and \Ii\ inside the Hall bar.

We find, that the overall current flow is chiral, in the sense that the current flows in opposite directions along the opposite edges of a Hall bar, while the imposed current is concentrated on the side of the Hall bar with the higher electrical potential, i.e., \Ii\ always vanishes along the edge with the lower potential. Our results are in agreement with recent experimental observations \cite{Sirt2025b}, but contradict the claim of Ref.\ \onlinecite{Guven2003}, that \Ii\ would flow in the same direction along both edges of the Hall bar.

\section{Screening Theory}\label{sec:screening_theory}
Exposing a 2DES to a high perpendicular magnetic field leads to the collaps of the density of states (DOS) into discrete LLs, the physical cause of the QHE. The Landauer-Büttiker picture of the QHE \cite{Buettiker1986,Buettiker1988}, neglects the Coulomb interaction between charge carriers. However, the energy gap between LLs would then lead to displacements of the electrons causing high local charges \cite{Chklovskii1992}. The screening theory fixes this problem by taking into account the direct Coulomb interaction between carriers \cite{Chklovskii1992,Chklovskii1993a,Fogler1994,Lier1994,Siddiki2003,Siddiki2004,Gerhardts2008}. This way, it accounts for the tendency of the carriers to minimize the free energy by redistribution (thereby avoiding high local charges) \cite{Chklovskii1992}. Nevertheless, in some regions of the 2DES, all LLs still end up to be either completely filled or empty, such that a local energy gap,
\begin{equation}
 \Delta E_\nu=\left\{\begin{array}{cl} g\mB B & \text{; for odd }\nu\\ \hbar\omegac-g\mB B & \text{; for even }\nu \end{array}\right.\,,
\end{equation}
remains, where $g$ is the Landee g-factor and \mB\ the Bohr magneton. For $\kb T\ll\Delta E_\nu$ and low enough disorder, this gap  locally suppresses scattering of electrons because of a lack of accessible unoccupied states. In a sense, the electrons behave like an incompressible liquid in these regions, which are therefore called incompressible strips (ICSs). These are the regions of suppressed scattering already mentioned in \sect{sec:introduction}. In more detail, the screening theory proposes a fragmentation of the Hall bar in compressible perfectly screened regions, separated by the unscreened ICSs \cite{Siddiki2003,Siddiki2004,Gerhardts2008}. Inside a compressible region, one partly filled LL is pinned to the locally constant chemical potential $\mu(y)$, hence the energies of all LLs, $\ve_\nu$, are also constant. Consequently, inside the unscreened ICSs, both, $\mu(y)$ and $\ve_\nu(y)$ vary, while the filling factor is constant.

The early formulations of the screening theory were content with analytical and first self-consistent approximations of the distribution of the electron density based on the Poisson equation \cite{Chklovskii1992} and using the Thomas-Fermi approximation \cite{Chklovskii1993a}. Later, the idea was refined by applying a self-consistent non-linear Hartree-type calculation and accounting for the quantum mechanical wave properties of the electrons by means of a Gaussian broadening of the LLs \cite{Siddiki2003,Siddiki2004}. Recently, (for high magnetic fields) the electrostatic results where confirmed for low integer filling factors within a full self-consistent approach of the quantum-electrostatic problem without using the Thomas-Fermi approximation \cite{Armagnat2020}.

\subsection*{Electrostatics of the QHE}

\begin{figure}[t]
\includegraphics[width=1\columnwidth ]{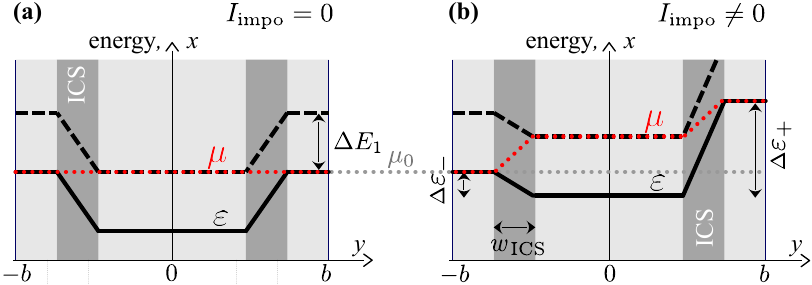}
\caption{Qualitative sketches of the predictions of the screening theory for the first plateau ($\nu=1$) of the QHE featuring the case of two ICSs near the edges of the Hall bar at $y=\mp b$. In (a), we assume $\Ii=0$, such that $\mu=\mu_0$ is constant. In (b), we illustrate the effect of $\Ii\ne0$ in the linear approximation, cf.\ main text. Energy $\ve(y)$ of the first LL (solid line), the second LL (dashed line), and the chemical potential $\mu(y)$ (dotted red line). Only inside the ICSs (dark gray background) $d\ve/dy\ne0$ and $d\mu/dy\ne0$.
}
\label{fig:ICSs}
\end{figure}
We consider a Hall bar with edges at $y=\pm b$, defined for the carrier density $\ns>0$ for $-b<y<b$. In \fig{fig:ICSs}{}, (assuming a symmetric confinement potential) we sketch the first two LLs (solid black line for the first and dashed black line for the second LL) as they are predicted by the screening theory near the low magnetic field end of the first plateau of the QHE. The energy gap between the first two LLs is $\Delta E_1=g\mu_\text{B}B$. In the considered limit, two narrow ICSs with filling factor $\nu=1$ form near the edges of the Hall bar. The energy of the LLs varies within the ICSs only. A current, \Ii, can be imposed between source (S) and drain (D) contacts, ideally positioned far away (at $x=\mp\infty$ along the vertical axis in the sketches in \fig{fig:ICSs}{}). In \fig{fig:ICSs}{a} we depict the equilibrium situation for $\Ii=0$ and constant $\mu_0(y)$, while in \fig{fig:ICSs}{b} we show a linear approximation for $\Ii\ne0$. For our example of two ICSs with $\nu=1$, the energy variation of the LLs across the two ICSs is $\Delta\ve_\pm=\pm\Delta E_1+\Delta \mu_\pm$, where $\mu_\pm$ correspond to the individual variations of the Hall potential in the two ICSs for $y\gtrless0$ with $\Delta\mu_++\Delta\mu_-=-e\vh$. Within the compressible strips between the edges and the ICSs, the first LL is pinned to the chemical potential ($\nu<1$). Within the compressible bulk region between the ICSs, the second LL is pinned to the chemical potential and partly occupied ($1<\nu<2$). In \sect{sec:current} below, we show that the current flows inside the ICSs. In our example in \fig{fig:ICSs}{}, it is carried by a single (spin polarized) LL.

For simplicity, we neglect many body correlations, interference effects or details at the interface between compressible regions and the ICSs. Further, we consider \Ii\ to be small compared to the break-down current \cite{Panos2014}, such that a quantized plateau and ICSs can exist, and assume an idealized 2DES at zero temperature without disorder and a symmetric confinement potential. Moderate disorder \cite{Siddiki2007} as well as asymmetric confinement \cite{Siddiki2010} would lead to corrections of the current distribution and influence the extent of the plateaus as well as the breakdown of the QHE. However, such influences will not affect the main findings of this article.

\subsection*{Geometry of the ICSs -- non-linear effects and linear approximation}
In \fig{fig:ICSs}{}, we assumed that the width, $w_\text{ICS}$, of the ICSs is independent of $\Ii$. This corresponds to the linear approximation of the screening theory for a symmetric Hall bar. A more accurate examination reveals a non-linear relation between the imposed current and the induced change of the electrostatic potential, which can be accounted for in a self-consistent numerical calculation \cite{Gerhardts2008}. The non-linear dynamics of the screening is already present in a simple electrostatic approximation considering the local energy balance. It considers the Coulomb energy, $E_\text{C}$, caused by charge separation across an ICS, which occurs for a constant filling factor inside of the ICSs. In steady state, the Coulomb energy equals the local energy variation of the LLs, $E_\text{C}=\Delta\ve_\pm$, which allows us to predict the width of an ICS, $w_\text{ICS}$. As long as the gradient of the (unscreened) confinement potential is constant within an ICS, $E_\text{C}\propto w_\text{ICS}^2$ is a reasonable approximation and we find $w_\pm\propto\sqrt{|\Delta \ve_\pm}|$ \cite{Chklovskii1992}. For $\Ii\ne0$ this result implies $w_+>w_-$, because $|\Delta\ve_+|-|\Delta\ve_-|=|e\vh|$, cf.\ \fig{fig:ICSs}{}. In effect, as we increase $|\Ii|$, the width $w_\text{ICS}$ of the ICS on the high potential side of the Hall bar is increased, while the width of the ICS on the low potential side of the Hall bar it is reduced. To keep our discussion simple, in \fig{fig:ICSs}{} we stick to the linear approximation, which corresponds to the first loop of the self-consistent iterations and  results in constant $w_\text{ICS}$ \cite{Gerhardts2008}.

\begin{figure}[t]
\includegraphics[width=1\columnwidth ]{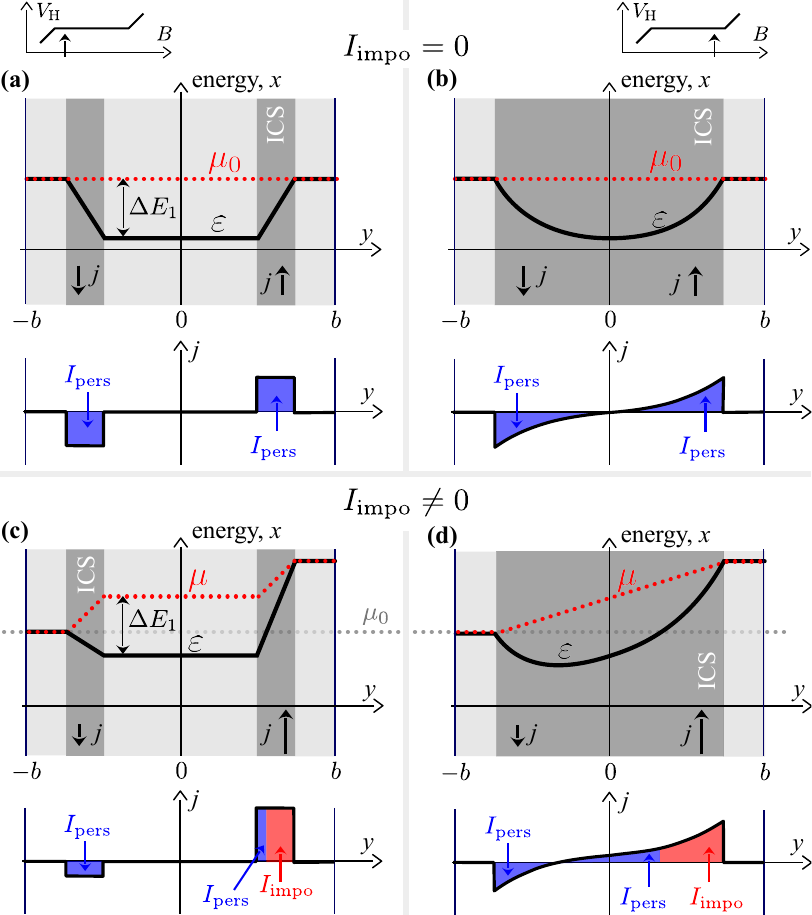}
\caption{Extension of Fig.\ \ref{fig:ICSs} featuring the current density for two different magnetic fields along the first plateau, as depicted in the insets at the upper end of the figure. Panels (a) and (c) on the left hand side illustrate the case of two separate ICSs, which occurs for $B$ near the low magnetic field end of the first plateau. Panels (b) and (d) on the right show one single ICS extending through most of the bulk of the Hall bar, which occurs for $B$ near the high magnetic field end of the first plateau. Shown are the energy $\ve(y)$ of the first LL (solid line), the chemical potential $\mu(y)$ (red dotted line), the current density directions along $\pm x$ inside the ICSs (dark gray background) and, in the line plots the actual current densities $j(y)$ in arbitrary units. Negative currents (flowing in the $-x$-direction) correspond to the integral $\int_{j<0}dy j$, positive currents to $\int_{j>0}dy j$. Regions with $\ve<\mu_0$ contribute to \Ip, regions with $\ve\ge\mu_0$ to \Ii. Finite \Ii\ causes \Ip\ to decrease.
}
\label{fig:electrostatics}
\end{figure}

Due to the curvature of the confinement potential, the geometry of the ICS also depends on the magnetic field: The carrier density and the filling factor increase with increasing distance from the sample edge. At the same time, at a given position inside the Hall bar, the filling factor decreases as $B$ is increased. Because, the integer filling factor has to remain constant inside the ICSs, as $B$ is increased each ICS will move from the edge towards the center of the Hall bar \cite{Chklovskii1992}. At the same time, its width increases because the absolute value of gradient of the confinement potential becomes smaller towards the center of the Hall bar. As $B$ is increased along a plateau of the QHE, this dynamics yields a transition from narrow edge ICSs to a wide bulk ICS towards the high magnetic field end of a plateau \cite{Siddiki2004}. In \fig{fig:electrostatics}{} we compare the extreme limits, narrow ICSs near the edges of the Hall bar in the left column versus a single bulk ICS in the right column. The bulk currents measured in Ref.\ \onlinecite{Sirt2025a} and the variations of $\mu(y)$ reported in \cite{Weitz2000} confirm the predicted \cite{Siddiki2004} formation of a bulk ICS. Note that, considering the non-linear screening behavior quantitatively would require self-consistent numerical calculations as done in Ref.\ \onlinecite{Siddiki2004}. While this would modify the detailed current distribution, the prediction of the following calculation remains otherwise valid.

\section{Current density distribution for the QHE}\label{sec:current}

\subsection*{Previous discussion of where the current flows}
The current density distribution for the case of a quantized Hall resistance became the subject of controversial discussions shortly after the discovery of the QHE, which are still continuing \cite{Laughlin1981,Halperin1982,Buettiker1988,Chklovskii1993a,Wexler1994,Geller1994,Komiyama1996,Guven2003,Gerhardts2020}. We provide only a few historical steps relevant for our discussion without claiming completeness. The common non-interacting Landauer-B\"uttiker picture assumes current flow within compressible one-dimensional edge channels \cite{Buettiker1988}. Taking into account the Coulomb interaction of the carriers,  Chklovskii et al.\ predicted the formation of ICSs \cite{Chklovskii1992} and then realized, calculating the ballistic current through a narrow gate defined channel, that for a quantized Hall resistance the current flows inside the ICSs \cite{Chklovskii1993a}. Gerhardts and collaborators used a local version of Ohm's law assuming that the imposed current density, $\ji(y)$, is proportional to the gradient of the chemical potential, $d\mu/dy$, which they calculated based on the screening theory \cite{Guven2003}. Since $d\mu/dy=0$ within compressible regions, they confirmed that the current flows inside the ICSs. However, by assuming $\ji\propto |d\mu/dy|$ in their calculations, they completely neglected the persistent current density, $\jp$, which led them to the prediction that the imposed current flows unidirectionally through all existing ICSs, i.e., in the same direction on both sides of the Hall bar \cite{Guven2003}. Another approach predicting \ji\ uses the continuity equation and finds $\ji\ne0$ inside compressible regions \cite{Armagnat2020}. We find this result questionable, mainly because it is based on calculating the linear response conductivity in the limit $\Ii\sim0$ by subtracting \jp\ from $\jp+\ji$ while $\jp\gg\ji$. However, for a finite \Ii\ current flow inside the ICSs becomes evident, once we accept that the gradient of the energy of a LL results in a drift velocity of the carriers within that LL, i.e., the local electric field drives current.

\subsection*{Model for the current density}
In the present article, we acknowledge in agreement with Refs.\ \onlinecite{Chklovskii1993a,Guven2003,Gerhardts2008,Gerhardts2020} that the current flows inside the ICSs. However, instead of focusing on the gradient of the chemical potential, which was used in Ref.\ \onlinecite{Guven2003} to predict \ji, we go back to the original approach by Chklovskii et al.\ \cite{Chklovskii1993a} and consider the complete current density $j=\ji+\jp$, which is carried by the fully occupied LLs inside the ICSs. We emphasize, that inside the unscreened ICSs the electrons of a LL are driven by the local electric field, which is identical to the gradient of the LL energy, $\vec E = -\vec\nabla \ve/e$ and differs from the gradient of the chemical potential.

The current density carried by an individual LL in the absence of scattering, i.e., inside an ICS, follows from fundamental relations: it is given by $\vec j_\text{LL} = -e\nl \vv$ with the velocity $\vv= \vec E\times\vec B /B^2$ of an electron and the carrier density $\nl=B\,e/h=\ns/\nu$ of the LL. Inside an ICSs the local filling factor, $\nu$, is a constant integer. For our Hall bar geometry we have $\vec E=-\frac{1}{e}\frac{d}{dy}(0,\ve,0)^\text{T}$, such that for $\vec B=(0,0,B)^\text{T}$ inside an ICS a current with density $\vec j=(j,0,0)^\text{T}$ flows in the $x$-direction with
\begin{equation}\label{current}
j=\nu j_\text{LL}=\pm\nu\frac{e}{h}\,\frac{d\varepsilon}{dy}\,.
\end{equation}
For all practical cases, the confinement potential of the Hall bar has opposite slopes along its two edges, cf.\ \fig{fig:electrostatics}{}. Therefore, $\vec E$ points in opposite directions and $j$ is positive on one side of the Hall bar and negative on its other side.

In \fig{fig:electrostatics}{}, we qualitatively sketch our prediction for $\nu=1$ based on \eq{current} and the screening theory \cite{Siddiki2004}. The four panels show for four different characteristic situations $\ve(y)\equiv\ve_1(y)$ (solid lines), $\mu(y)$ (dotted red lines) and the related $j(y)$ (lower panels) across a Hall bar. ICSs are indicated by a dark gray background, compressible regions by light gray. Only within the unscreened ICSs the LL energy, $\ve(y)$, varies and $j\ne0$. In Figs.\ \ref{fig:electrostatics}{a} and \ref{fig:electrostatics}{b}, we consider the equilibrium with $\Ii=0$, while in Figs.\ \ref{fig:electrostatics}{c} and \ref{fig:electrostatics}{d} we added a finite $\Ii\ne0$.  In Figs.\ \ref{fig:electrostatics}{a} and \ref{fig:electrostatics}{c}, we depict the case of two ICSs near the edges of the Hall bar, while in Figs.\ \ref{fig:electrostatics}{b} and \ref{fig:electrostatics}{d}, we sketch the case of a single bulk ICS. The screening theory predicts these two cases to occur at the two ends of the plateau of a quantized Hall resistance as indicated in the insets above Figs.\ \ref{fig:electrostatics}{a} and \ref{fig:electrostatics}{b} \cite{Siddiki2004}.

In the contacts, the confinement potential is screened, if they are compressible, and no persistent current flows. Hence, in order to comply with Kirchhoff's junction rule (or particle conservation), \Ip\ has to form a closed loop inside the Hall bar. Clearly, its density fulfills $\int_{-b}^b dy \jp=0$. Hence, we can define \Ip\ by dividing this integral into two regions: $\Ip\equiv-\int_{-b}^{y(j=0)} dy \jp=\int_{y(j=0)}^b dy \jp$, which has the identical absolute value on both sides of the Hall bar. In contrast to \Ip, the imposed current enters the Hall bar and leaves it through the source and drain contacts. Therefore, \Ii\ corresponds to the integral of the complete current density over the full width of the Hall bar, $\Ii=\int_{-b}^b dy j=\int_{-b}^b dy \ji=|\vh|\nu e^2/h$, where for integer filling factor $\nu=1,2,3,\dots$, the term $h/(\nu e^2)$ is the quantized Hall resistance.

The current density \ji\ generates the Hall potential between the two edges, which corresponds to the change of the chemical potential according to $\vh(y)=\left[\mu(y)-\mu_0\right]/e$, cf.\ Figs.\ \ref{fig:electrostatics}{c} and \ref{fig:electrostatics}{d}. The quantity $\vh(y)$ is measured in the scanning probe experiment in Ref.\ \onlinecite{Weitz2000}, if $\mu_0$ is the reference ground potential. In our sketches we chose the chemical potential at the drain contact as this reference, $\mu(-b)=\mu_0$. In equilibrium, i.e., for $\Ii=0$, the chemical potential is constant, $\mu(y)=\mu_0$. For $\Ii\ne0$, $\mu(y)\ge\mu_0$ increases monotonously between the edges, while it always remains constant within the compressible regions. An increase of $\mu(y)$ corresponds to an accumulation of a negative charge. Therefore, $\ve(y)$ increases accordingly (compare the upper panels for $\Ii=0$ with the lower panels for $\Ii\ne0$ in \fig{fig:electrostatics}{}). Importantly, the increase of $\ve(y)$ yields a decrease of $j$ for $-b<y\lesssim0$ (the low potential side of the Hall bar) and an increase of $j$ for $0\lesssim y<b$ (the high potential side). Because $\int_{-b}^b dy \jp=0$, the decrease of $j$ on the low potential side corresponds to a decrease of \Ip. The increase of $j$ on the high potential side then points to a combination of the established decrease of \Ip\ together with \Ii\ flowing there.

Within our linear approximation, see \sect{sec:screening_theory}, for $e\vh\ll2\Delta E_\nu$ we find an opposite direction of the current flow on the two sides of the Hall bar, i.e., a chiral current flow. The case $e\vh\sim2\Delta E_\nu$, which is clearly beyond the scope of our approximation, would result in a break down of the ICSs. However, as long as an ICS exists on the low potential side, we expect $d\ve/dy$ to remain $\le0$ there, supported by numerical calculations \cite{Panos2014}. We conclude: whenever the Hall resistance is quantized (ICSs exist), the current flow is chiral.

\begin{figure}[t]
\includegraphics[width=1\columnwidth]{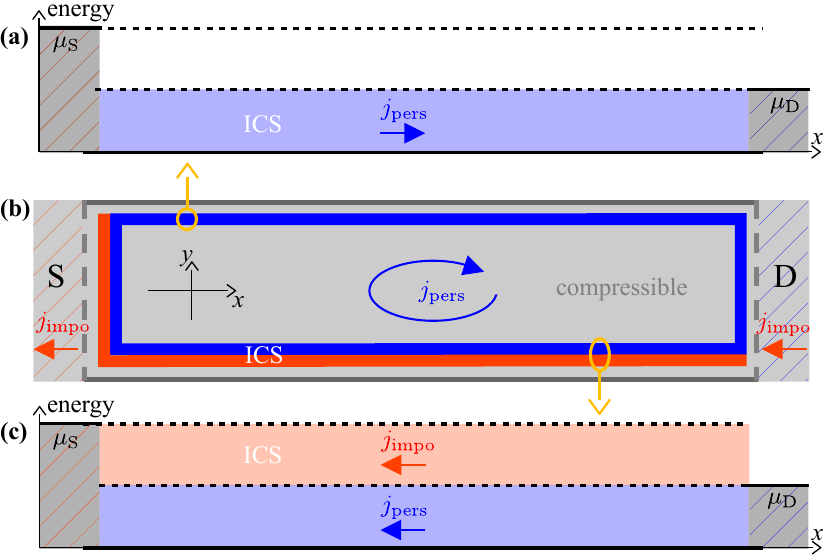}
\caption{Illustration of geometry in (b) of the ICS corresponding to the non-equilibrium situation with $\Ii\ne0$ of Fig.\ \ref{fig:electrostatics}(c); (a) and (c) are energy diagrams of the ICS along the upper and lower edges, repsecively. States contributing to \jp\ are indicated in blue and those of \ji\ in red. The persistent current describes a closed path inside the Hall bar as all contributing states have energies below the chemical potentials of both leads, $E<\mus,\mud$. The imposed current is carried by states with energies $\mus\ge E>\mud$. Its electrons enter the Hall bar from the source contact, move along its high potential (lower) edge and leave it into empty states of the drain contact.
}
\label{fig:chiral_current2}
\end{figure}

The illustration in \fig{fig:chiral_current2}{},
offers an alternative point of view for the case of two ICSs in the vicinity of the edges. In panel (b), we sketch a minimalistic model of a Hall bar. It consists of two edges connecting the source (S) and drain (D) contacts fixed to the chemical potentials $\mus>\mud$. The persistent current flows in a closed loop inside the ICS (blue). The imposed current flows through the ICS from drain to source (red). The corresponding electrons moving from source to drain pin the chemical potential of the lower edge to \mus. Since \ji\ flows entirely along the lower edge of the Hall bar, its upper edge remains at \mud. Figures \ref{fig:chiral_current2} (a) and (c) illustrate the chemical potentials of the contacts and the energy range of the current carrying (occupied) states inside the ICS along the upper versus lower edge, respectively. Shaded regions indicate occupied states (at $T=0$), gray inside the contacts, blue inside the ICS at energies $E<\mud$, and red inside the ICS at energies $\mud<E\le\mus$. Electrons with $E\le\mud$ (blue) contribute to \jp, electrons with $E>\mud$ (red) contribute to \ji. Clearly, \jp\ follows a closed path inside the Hall bar as for $E\le\mud$ no unoccupied states are available in the leads. In contrast \ji\ is carried by electrons emitted from the source contact, which then leave the Hall bar into the drain contact, cf.\ \fig{fig:chiral_current2}{c}. Reflecting the chiral nature of the ICSs, predifined by the local gradients of the LL energies, cf.\ \fig{fig:electrostatics}{b}, \ji\ flows entirely inside the ICS along the high potential (lower) edge.

\section{Summary}
In summary, by applying the screening theory and considering the complete current density $j=\ji+\jp$ inside the Hall bar, we find that for a quantized Hall resistance
\begin{enumerate}[label=(\roman*)]
 \item the current flows inside ICSs,
 \item the current flow is chiral,
 \item the current density consist of a persistent and an imposed contribution, $j=\jp+\ji$,
 \item $\Ip$ is maximal for $\Ii=0$ and decreases as $|\Ii|$ is increased,
 \item \Ii\ flows entirely on the high potential side of the Hall bar.
\end{enumerate}
Finally, we note that the occurrence of bulk current is not in conflict with chiral current flow as long as scattering of the carriers is suppressed inside the bulk ICS, cf.\ \fig{fig:electrostatics}{}.

\section*{Data availability statement}

This article contains no experimental or theoretical data. Therefore, no data can be made publicly available.

\section*{Acknowledgement}

This work was funded by the Deutsche Forschungsgemeinschaft (DFG, German Research Foundation) -- 218453298.

\section*{Contributions of the authors}

S.\,L.\ wrote the article. S.\,S.\ contributed in many fruitful discussions and carefully read the article.

\section*{References}
%

\end{document}